\documentclass[letterpaper, 10 pt, conference]{ieeeconf}  

\IEEEoverridecommandlockouts                              
\usepackage{graphics} 
\usepackage{epsfig} 
\usepackage{amsfonts}       
\usepackage{nicefrac}       
\usepackage{microtype}      
\usepackage{xcolor}         
\usepackage{graphicx,wrapfig}

\usepackage{xcolor}
\usepackage{tikz}
\usepackage[hidelinks]{hyperref}

\usepackage{amssymb,amsmath}
\usepackage{multirow,multicol}
\usepackage{algorithm,algorithmic}
\usepackage{cite}

\newtheorem{example}{Example}

\definecolor{matplotblue}{RGB}{0, 0, 225} 

\newcommand{\bluecircle}{\begin{tikzpicture}[baseline=(current bounding box.base)]
    \fill[matplotblue] (0,.1) circle (0.12);
\end{tikzpicture}}

\newcommand{\bluetriangle}{\begin{tikzpicture}[baseline=(current bounding box.base)]
    \fill[matplotblue] (0,0) -- (.125,.2165) -- (.25,0) -- cycle;
\end{tikzpicture}}

\newcommand{\red}[1]{{\color{red} \noindent #1}}

\title{\LARGE \bf Domain Adaptive Safety Filters via Deep Operator Learning}

\author{Lakshmideepakreddy Manda$^1$, Shaoru Chen$^2$, Mahyar Fazlyab$^1$
\thanks{$^{1}$Department of Electrical and Computer Engineering,
        Johns Hopkins University, 3400 N Charles St, Baltimore, MD, 21218.
        {\tt\small lmanda1@jhu.edu}}%
\thanks{$^{2}$Microsoft Research, 300 Lafayette Street, New York, NY, 10027.
        {\tt\small shaoruchen@microsoft.com}}%
\thanks{$^{1}$Department of Electrical and Computer Engineering,
        Johns Hopkins University, 3400 N Charles St, Baltimore, MD, 21218.
        {\tt\small mahyarfazlyab@jhu.edu}}%
}

\begin{document}
\maketitle
\thispagestyle{empty}
\pagestyle{empty}

\begin{abstract}
Learning-based approaches for constructing Control Barrier Functions (CBFs) are increasingly being explored for safety-critical control systems. However, these methods typically require complete retraining when applied to unseen environments, limiting their adaptability. To address this, we propose a self-supervised deep operator learning framework that learns the mapping from environmental parameters to the corresponding CBF, rather than learning the CBF directly. Our approach leverages the residual of a parametric Partial Differential Equation (PDE), where the solution defines a parametric CBF approximating the maximal control invariant set. This framework accommodates complex safety constraints, higher relative degrees, and actuation limits. We demonstrate the effectiveness of the method through numerical experiments on navigation tasks involving dynamic obstacles.
\end{abstract}

\section{introduction}
\label{sec:intro}
A crucial criterion for control design in safety-critical domains is to ensure that the system's trajectories will always remain in a safe region. One popular approach to enforce safety constraints is through control barrier functions (CBFs), see e.g.,~\cite{ames2019control}. CBF-based safety filters minimally correct a legacy controller to ensure constraint satisfaction. Specifically, for continuous-time control-affine systems, this involves deploying a convex quadratic program, referred to as CBF-QP, that can be solved efficiently in real-time.  

Despite their simplicity, there is no general methodology for constructing CBFs for arbitrary nonlinear systems. The main challenges in CBF synthesis include bounded actuation~\cite{liu2023safe}, high relative degree~\cite{nguyen2016exponential, xiao2021high}, complex safety specifications~\cite{molnar2023composing}, and the inherent conservatism of CBFs~\cite{chen2024learning}. To address these issues, machine learning techniques have increasingly been employed to improve the synthesis of CBFs for more complex systems and environments~\cite{dawson2023safe, robey2020learning, dawson2022safe, qin2021learning}. However, a key limitation of data-driven approaches, which motivates the current work, is that CBFs learned in one environment may become invalid in another. This environment dependence contrasts with Model Predictive Control (MPC), which can adapt to changing environments.

\subsubsection*{Our Contribution} Motivated by the challenges in CBF synthesis and adaptation to changing environments, in this work, we propose a deep operator learning framework that learns the mapping between the environment and the CBF, as opposed to learning the CBF itself. Following \cite{chen2024learning}, our methodology starts with a Hamilton-Jacobi (HJ) type parametric Partial Differential Equation (PDE), in which the finite-dimensional parameters characterize the environment (e.g., the location of obstacles), and the unique solution of the PDE yields a CBF that approximates the maximal control invariant set. Using the residual of this PDE as a self-supervised training loss function, we learn the solution operator of the PDE, dubbed the \emph{CBF operator}, using a continuously differentiable Multi-Layer Perceptron (MLP) architecture. The resulting CBF  and the corresponding safety filter are an explicit function of the environment parameters, enabling it to adapt to novel or changing environments during deployment. The overall framework takes bounded actuation, high relative degree, complex safety specifications, and conservatism of CBF into account. We demonstrate the efficacy of our method on various safe control tasks with moving obstacles.

\subsection{Literature Review}
\subsubsection{CBF synthesis}
Several key challenges in constructing a CBF have been investigated in the literature. For multiple safety constraints that are possibly associated with logical operations, Glotfelter et al. \cite{glotfelter2017nonsmooth} compose multiple CBFs accordingly through the non-smooth min/max operators. In \cite{lindemann2018control}, a smooth composition of logical operations on constraints were introduced and later extended to handle actuation and state constraints simultaneously  \cite{rabiee2023softmin, rabiee2023softminmax} using integral CBFs \cite{ames2020integral}.  Molnar et al. \cite{molnar2023composing} propose an
algorithmic way to create a single smooth CBF arbitrarily composing both min and max operators. Such smooth bounds have been used in changing environments~\cite{safari2023time}. Notably, \cite{breeden2023compositions} ensures the input-constrained feasibility of the CBF condition while composing multiple CBFs. 

High-order CBF (HOCBF)~\cite{xiao2021high} and exponential CBF~\cite{nguyen2016exponential} systematically approach CBF construction when the safety constraints have high relative degree. However, controlling the conservatism of these approaches is a challenge. To reduce conservatism or improve the performance of HOCBF, learning frameworks have emerged~\cite{xiao2023barriernet, xiao2023learning, ma2022learning} allowing tuning of the class $\mathcal{K}$ functions used in the CBF condition. 

Learning frameworks~\cite{dawson2023safe, robey2020learning, dawson2022safe, qin2021learning} building on~\cite{djeridane2006neural} have tackled the challenge of limited actuation. Liu et al.~\cite{liu2023safe} explicitly consider input constraints in learning a CBF by finding counterexamples on the $0$-level set of the CBF for training, while Dai et al.~\cite{dai2023data} propose a data-efficient, prioritized sampling method. 

In~\cite{nakamura2021adaptive}, an adaptive sampling method for training HJ NN models favoring sharp gradient regions is proposed. Drawing tools from reachability analysis, the recent work~\cite{so2023train} iteratively expands the volume of the control invariant set by learning the policy value function and improving the performance through policy iteration. Similarly, Dai et al.~\cite{dai2023learning} expand conservative hand-crafted CBFs by learning on unsafe and safe trajectories. In~\cite{tan2023your}, an actor-critic reinforcement learning framework is applied to learn a CBF value function. Qin et al.~\cite{qin2022sablas} learn a control policy and CBF together for black-box constraints and dynamics. We exploit the connection between CBFs and Hamilton-Jacobi (HJ) reachability analysis investigated in~\cite{choi2021robust, tonkens2022refining} for CBF learning in this work. 

Unlike any of the mentioned learning approaches, our CBF learning approach considers both bounded actuation and complex safety constraints while aiming to learn the maximal control invariant set. Importantly, our method considers changing environments by learning a NN CBF operator.

\red{}

\subsubsection{Adaptive CBF} 
Adaptive CBFs (aCBF) \cite{taylor2020adaptive} considers parametric model uncertainty and proposes an adaptive CBF that is robust against a bounded uncertainty set. Robust adaptive CBFs~\cite{lopez2020robust} improve the conservatism of aCBF and introduce set membership identification to reduce modeling errors. In the presence of system disturbances and measurement noises, Takano et al.~\cite{takano2018robust} apply a robust CBF formulation for safety filter design. Xiao et al.~\cite{xiao2021adaptive} propose an adaptive high-order CBF that adapts to time-varying control inputs and noises in the dynamics. Close to our work are~\cite{molnar2022safety} which formulates an environmental CBF that incorporates the environmental parameters into CBF parameterization, and \cite{ma2022learning} which focuses on tuning the class $\mathcal{K}$ function in CBF to achieve good performance across environments. Different from~\cite{molnar2022safety}, which focuses on handling delay inputs and case studies assuming the availability of the environmental dynamics, and~\cite{ma2022learning}, which relies on the exponential CBF formulation, our work aims to provide a holistic, automated learning method for CBF construction across various environments with complex safety specifications. 

\subsection{Notation}
An extended class $\mathcal{K}$ function is a function $\alpha:(-b, a) \mapsto \mathbb{R}$ for some $a, b > 0$ that is strictly increasing and satisfies $\alpha(0) = 0$. We denote $L_r^{+}(h) = \{x \in \mathbb{R}^n \mid h(x) \geq r \}$ as the $r$-superlevel set of the function $h$. The positive and negative parts of a number $a \in \mathbb{R}$ are denoted by $(a)_{+}=\max(a,0)$ and $(a)_{-}=\min(a,0)$, respectively. We denote the class of $k$-times continuously differentiable functions by $C^k$ ($k \geq 0$). We denote the perspective of the log-sum-exp (LSE) function by $\text{LSE}(y;\beta)=\frac{1}{\beta}\log(\sum_{i=1}^M\exp(\beta y_i))$ for $\beta>0$.

\section{Background}
\label{sec:prelim}
Consider a continuous time control-affine system:
\begin{align} \label{eq:dynamics}
    \dot{x} = f(x) + g(x) u, \quad u \in \mathcal{U}, 
\end{align}
where $f\colon \mathbb{R}^n \to \mathbb{R}^n$, $g \colon \mathbb{R}^n \to \mathbb{R}^{n \times m}$ are locally Lipschitz continuous, $x \in \mathcal{D} \subseteq \mathbb{R}^{n}$ is the state with domain $\mathcal{D}$, $u$ is the control input, and $\mathcal{U} \subseteq \mathbb{R}^m$ is the control input constraint set, which we assume to be a convex polyhedron. We denote the solution of \eqref{eq:dynamics} at time $t \geq 0$ by $x(t)$. 

Given a set $\mathcal{X} \subseteq \mathcal{D}$ that represents a safe subset of the state space, the general objective of safe control design is to find a control law $\pi(x)$ that renders $\mathcal{X}$ forward invariant under the closed-loop dynamics $\dot{x} = f(x) + g(x) \pi(x)$, i.e., if $x(t_0) \in \mathcal{X}$ for some $t_0 \geq 0$, then $x(t) \in \mathcal{X}$ for all $t \geq t_0$. In this paper, we assume that the safe set can be characterized by the $0$-superlevel set of a continuous function $c(\cdot) \in C^0$,
\begin{align}
    \mathcal{X} = L_0^{+}(c)= \{x \mid c(x) \geq 0\}.
\end{align}

\subsection{Control Barrier Functions (CBFs)}
A continuously differentiable function $h \colon \mathbb{R}^n \to \mathbb{R}$ is a CBF if the following conditions hold: \emph{(i)} The 0-super level set of $h$ is contained in $\mathcal{X}$, i.e,
\begin{align}\label{eq: cbf condition 1}
        L_0^{+}(h) = \{ x\mid h(x) \geq 0\} \subseteq \mathcal{X}.
    \end{align}
    \emph{(ii)} A Lyapunov-like condition on the whole domain: 
    \begin{align}\label{eq: cbf condition 2}
        \sup_{u \in \mathcal{U}} \{\nabla h(x)^\top (f(x)+g(x)u)\} \!+\! \alpha(h(x)) \geq 0 \ \forall x \in \mathcal{D},
    \end{align}
    where $\alpha$ is an extended class $\mathcal{K}$ function, typically chosen to be linear, $\alpha(h(x)) = \gamma h(x), \ \gamma>0$.  If the system has unbounded actuation, $u \in \mathbb{R}^{m}$, \eqref{eq: cbf condition 2} becomes equivalent to
\begin{align}
    \nabla h(x)^\top g(x) = 0 \implies \nabla h(x)^\top f(x) + \alpha(h(x))\geq 0.
\end{align}
The above conditions together ensure that $L_0^{+}(h)$ is a control invariant subset of $\mathcal{X}$. 

We denote the class of all admissible CBFs by $\mathcal{H}$:
\begin{align}
    \mathcal{H} = \{h \in C^1 \mid h \quad \text{satisfies } \eqref{eq: cbf condition 1} \text{ and } \eqref{eq: cbf condition 2} \}. 
\end{align}
The CBF can be used as a safety filter, framed as a convex QP, to ensure safety for a given baseline control policy $u_r(x)$,
\begin{alignat}{2} \label{eq: safety filter CBF}
        &u_{s}(x):= \ &&\text{argmin}_{u \in \mathcal{U}}  \ \lVert u - u_r(x) \rVert^2 
        \\
        & && \text{s.t.}  \ \nabla h(x)^\top \left(f(x) +g(x) u \right) + \alpha(h(x)) \geq 0, \notag
\end{alignat}

\subsection{Neural Network-based CBF}
%
The function \( c(\cdot) \), which defines the boundary of the safe set \(\mathcal{X}\), often fails to qualify as a CBF, either due to a lack of differentiability or because it does not satisfy the invariance condition \eqref{eq: cbf condition 2}. Thus, we must construct a CBF, either on a case-by-case basis or using function approximation techniques. 

Specifically, let \( h_{\theta}(x): \mathbb{R}^n \rightarrow \mathbb{R} \) be the parameterized function approximator. The parameters $\theta$ can be found by solving the following self-supervise risk minimization problem, in which the violation of both conditions \eqref{eq: cbf condition 1} and \eqref{eq: cbf condition 2} is penalized:
\begin{align} \label{eq: CBF risk}
    \min_{\theta} \mathbb{E}_{x \sim \rho} [{(-\mathcal{I}_{\gamma}(h_{\theta}(x)))_{+}}^{2}+\lambda {\left(h_{\theta}(x)-c(x)\right)_{+}}^{2}], 
\end{align}
Here $\rho$ is typically a uniform distribution over $\mathcal{D}$ and 
\begin{equation} \label{eq:CBF condition function}
\begin{aligned}
\mathcal{I}_{\gamma}(h):= \sup_{u \in \mathcal{U}} \{\nabla h(x)^\top (f(x)+g(x)u)\} \!+\! \gamma h(x).
\end{aligned}
\end{equation}
This approach, however, focuses solely on satisfying the invariance conditions and does not incorporate notions of performance. Consequently, it may result in an overly conservative CBF in practice. 

\subsection{Performance-Oriented CBF via HJ PDE}

Ideally, we want to find a CBF $h$ whose zero super level set $L_0^+(h)$ has the maximal volume within the safe set $\mathcal{X}$. To achieve this, we can characterize the maximal control invariant set inside $\mathcal{X}$ through an HJ PDE. Consider the dynamics~\eqref{eq:dynamics},
\begin{equation} \label{eq:dynamics_t}
    \dot{x}(\tau) = f(x(\tau)) + g(x(\tau)) u(\tau), \ \tau \in [t, 0], \ x(t) = x,
\end{equation}
on a time interval $[t,0]$ where $t \leq 0$ is the initial time, and $x$ is the initial state. Let $\mathcal{U}_{[t,0]}$ be the set of Lebesgue measurable functions $u\colon [t,0] \to \mathcal{U}$, and $\psi(\tau):= \psi(\tau; x, t, u(\cdot)): [t, 0] \mapsto \mathbb{R}^n$ denote the unique solution of~\eqref{eq:dynamics_t} given $x$ and $u(\cdot) \in \mathcal{U}_{[t,0]}$. Given a bounded Lipschitz continuous function $c: \mathcal{D} \mapsto \mathbb{R}$, the viability kernel of $L_{0}^{+}(c) =\{x \mid c(x) \geq 0\}$ is defined as 

\begin{equation}
\begin{aligned}
    \mathcal{V}(t) \! := \! \{ x \in L_{0}^{+}(c) & \mid \exists u(\cdot) \in \mathcal{U}_{[0,t]}, \\
    & \text{ s.t. } \forall \tau \in [t, 0], \psi(\tau) \in L_{0}^{+}(c) \}.
\end{aligned}
\end{equation}
Taking $t \rightarrow -\infty$, the viability kernel gives us the \emph{maximal} control invariant set contained in $L_{0}^{+}(c)$~\cite{choi2021robust}. For $t \leq 0$, the viability kernel $\mathcal{V}(t)$ can be expressed as the super level set of the control barrier-value function $B_\gamma : \mathcal{D}\times (-\infty,0] \mapsto \mathbb{R}$ defined as 
\begin{equation} \label{eq:cbvf_def}
    B_\gamma(x,t):=\underset{u(\cdot) \in \mathcal{U}_{[t,0]}}{\max} \ \underset{\tau \in [t, 0]}{\min} e^{\gamma(\tau - t)} c(\psi(\tau)),
\end{equation}
where $\gamma \geq 0$. In other words, for each $\gamma \geq 0$ and all $t \leq 0$, we have $\{ x \mid B_\gamma(x, t) \geq 0\}  = \mathcal{V}(t)$ \cite{choi2021robust}. Furthermore, $B_\gamma$ is a unique Lipschitz continuous viscosity solution of the HJ PDE:
    \begin{equation} \label{eq:cbvf_pde}
    \begin{aligned}
        & \min   \{ c(x) - B_\gamma(x, t), \frac{\partial}{\partial t} B_\gamma(x,t) + \\
        & \max_{u \in \mathcal{U}} \nabla_x B_\gamma(x,t) ^T (f(x) + g(x)u) + \gamma B_\gamma(x,t) \} = 0, 
    \end{aligned}
    \end{equation}
with the boundary condition $B_\gamma(x,t) = c(x)$. We obtain the steady-state HJ PDE by taking $t \rightarrow -\infty$:
    \begin{equation} \label{eq:hj_pde}
    \begin{aligned}
        &\mathcal{N}_{\gamma}(c,B_{\gamma}):= \min  \{ c(x) - B_\gamma(x),  \\
        &\max_{u \in \mathcal{U}} \nabla_x B_\gamma(x) ^T (f(x) + g(x)u) + \gamma B_\gamma(x) \} = 0.
    \end{aligned}
\end{equation}
For a given input function $ c(\cdot)$ to this PDE, the unique solution $B_\gamma(\cdot)$, when differentiable everywhere, is a valid CBF, satisfying both conditions \eqref{eq: cbf condition 1} and \eqref{eq: cbf condition 2}. In the case where not fully differentiable, for certain representative cases, e.g., when $c(\cdot)$ is a signed distance function and the system dynamics is Lipschitz continuous, $B_\gamma(\cdot)$ will be differentiable almost everywhere \cite{wabersich_data-driven_2023}.

Building on this, we can adopt a similar approach to the previous risk minimization formulation \eqref{eq: CBF risk}. However, this time, we focus on minimizing the residual of the PDE \eqref{eq:hj_pde} 
\begin{align} \label{eq: HJ risk}
    \min_{\theta} \mathbb{E}_{x \sim \rho} [(\mathcal{N}_{\gamma}(c(x),h_{\theta}(x)))^2]. 
\end{align} An appealing feature of the above training loss function is that it admits an ``optimal'' CBF in the sense that $L_0^{+}(h)$ is approximately maximal.
\section{Main Idea: Domain Adaptation via Operator Learning}
\label{sec:operator}
The learned solution of problem \eqref{eq: HJ risk} will be valid only for a specific environment defined by the function $c(\cdot)$. As a result, it cannot be deployed in novel environments without retraining. This limits its generalization and usefulness in real-time applications.

Instead of retraining a new CBF for each different environment, we propose to learn the \emph{solution operator} of the PDE. Formally, given any $c(\cdot) \in C^0$ and the corresponding solution $h(\cdot) \in C^1$ that satisfies $\mathcal{N}_{\gamma}(c(x),h(x))=0 \ \forall x$, we define the solution operator $\mathcal{G} \colon C^0 \to C^1$ as 
\begin{align}
    h(x) = \mathcal{G}(c(x)).
\end{align}
$\mathcal{G}(\cdot)$ is a nonlinear operator that maps between infinite dimensional Banach spaces. To approximately learn this map, we assume that the input function $c(\cdot)$ is parameterized by a \emph{finite-dimensional} vector $e \in \mathbb{R}^{n_e}$ that represents the parameters of the environment, e.g., the location and sizes of obstacles. Under this assumption, we can abstract the safe set as $\mathcal{X} = \{x \mid c(x,e) \geq 0\}$. 

By viewing the environment parameters as ``virtual'' states without dynamics, i.e., $\dot{e}=0$, we define the augmented state as $\xi = (x,e)$, obeying the following dynamics,
\begin{align}
    \dot{\xi} = f^{\mathrm{aug}}(\xi)+ g^{\mathrm{aug}}(\xi) u,
\end{align}
where 
\begin{align}
  f^{\mathrm{aug}}(\xi) = \begin{bmatrix}
      f(x) \\ 0
  \end{bmatrix},  \quad  g^{\mathrm{aug}}(\xi) = \begin{bmatrix}
      g(x) \\ 0
  \end{bmatrix}.
\end{align}

Using this augmentation, our approach is to learn a neural CBF in the \emph{joint space} of the states and the environment parameters using a loss akin to \eqref{eq: HJ risk},
\begin{align} \label{eq: HJ augmented risk}
    \min_{\theta} \mathbb{E}_{\xi \sim \rho} [(\mathcal{N}_{\gamma}(c(\xi),h_{\theta}(\xi)))^2]. 
\end{align}
In the next subsection, we provide the details of the proposed training loss.

\subsection{Training Loss Function}

To train the CBF operator, $h_\theta(x, e)$, we adopt a similar procedure as~\cite{chen2024learning} which for each environment approximates the maximal control invariant set characterized by the HJ PDE~\eqref{eq:hj_pde} and enhances the feasibility of the safety filter with a CBF condition penalty term. We uniformly sample the environmental parameters $E = \{e_i\}_{i=1}^M$ and the states $X_e = \{x_e^{j}\}_{j=1}^N$ for each environment. Then, we construct the total dataset of joint states as $\Xi = \bigcup_{i=1}^M \{e_i\}\times X_{e_i}$, where $\times$ denotes the Cartesian product of sets. We train the CBF operator $h_\theta(x,e)$ using two loss functions $\mathcal{L}_{\text{HJ}}$ and $\mathcal{L}_{\text{CBF}}$. 

\subsubsection*{$\mathcal{L}_{\text{HJ}}$ for Maximal Control Invariant Set Approximation} 
This loss guides the CBF $h_\theta(\cdot, e)$ to approximate the maximal control invariant set for each environment
\begin{equation}\label{eq:phase_1_obj}
\begin{aligned}
    \ & \mathcal{L}_{\text{HJ}}(\theta) = \frac{1}{|\Xi|} \sum_{\xi\in \Xi} (\mathcal{N}_{\gamma}(c(\xi),h_{\theta}(\xi)))^2. 
\end{aligned}
\end{equation}
When the control input set $\mathcal{U}$ is polyhedral, the maximum over $u\in \mathcal{U}$ in the PDE is achieved at one of the vertices and can be obtained in closed form. 

\subsubsection*{$\mathcal{L}_{\text{CBF}}$ for Feasibility Enhancement} The HJ PDE is derived for the policy that achieves the maximal control invariant set; hence, the safe control set $\{ u \in \mathcal{U} \mid \nabla_x h_\theta(x, e)^\top (f(x) + g(x) u) + \gamma h_\theta(x, e) \geq 0\}$ at state $x$ can be small or even be a singleton. This tends to make the CBF-QP~\eqref{eq: safety filter CBF} infeasible in practice. Therefore, we also explicitly enforce the second CBF condition~\eqref{eq: cbf condition 2} with $\mathcal{L}_{\text{CBF}}$, 
\begin{equation} \label{eq:phase_2_obj}
\begin{aligned}
\ & \mathcal{L}_{\text{CBF}}(\theta) = \frac{1}{|\Xi|} \sum_{\xi\in \Xi} [-\mathcal{I}_{\gamma}(h_{\theta}(\xi))_+]^2,
\end{aligned}
\end{equation}
which improves the feasibility of the CBF-QP. We combine $\mathcal{L}_{\text{HJ}}$ and $\mathcal{L}_{\text{CBF}}$ into a single objective with scalarization hyperparameter $\lambda>0$,
\begin{equation} \label{eq:combined_loss}
\underset{\theta}{\text{min}}\; \mathcal{L}(\theta) = \mathcal{L}_{\text{HJ}}(\theta) + \lambda\mathcal{L}_{\text{CBF}}(\theta).
\end{equation}
We use standard first-order optimization methods for this objective. We note that the first CBF condition ~\eqref{eq: cbf condition 1}, $c(\xi)\geq h_{\theta}(\xi)$, does not appear explicitly in the above loss function. Our specific parameterization will guarantee this property. The details of this parameterization, which plays a crucial role in improving the learning process, will be discussed in the following subsection.

\subsection{CBF Parameterization}
Instead of learning the CBF operator \(h\) directly, we can learn the difference function \(c-h\). To formalize this, we introduce a parameterized CBF operator as
\begin{equation} \label{eq:nonsmooth parameterization}
    h_\theta(x, e) = c(x, e) - \delta_\theta(x, e),
\end{equation}
where \(\delta_\theta(\cdot, e): \mathbb{R}^{n + n_e} \mapsto \mathbb{R}\) is a non-negative valued function approximator. Importantly, by construction, we have $h_\theta(x, e) \leq c(x,e)$ which implies that the $0$-superlevel set of $h_\theta(\cdot, e)$ is always contained in the safe region $\mathcal{X}$ \emph{for all environments}. We parameterize the difference function \(\delta_\theta(x, e)\) as the following MLP,
\begin{equation}
\begin{aligned}
    z_0 &= [x^\top \ e^\top]^\top, \\
    z_{k+1} &= \sigma(W_k z_k + b_k), k = 0, \cdots, L-1, \\ 
    \delta_{\theta}(x, e) &= \sigma_+(W_L z_L + b_L), \\
\end{aligned}
\end{equation}
where $W_k, b_k$ denote the weights and bias of the $(k+1)$-th linear layer, $\sigma(\cdot)$ is a smooth activation, and $\sigma_+(\cdot)$ is a smooth non-negative valued activation such as the Softplus function. Note that $\delta_\theta(x,e)$ is a function of all the states and so is the CBF operator $h_\theta$. Therefore, we alleviate the issue of high relative degree~\cite{xiao2021high} through the parameterization of the MLP CBF operator.

However, the above parameterization is valid only when \( c(x,e) \) is continuously differentiable. In the next subsection, we address cases where \( c(x,e) \) is not differentiable or where the environment includes multiple safety constraints.

\subsection{Complex Enviroments}
The PDE \eqref{eq:hj_pde} takes a single constraint function $c(\cdot, \cdot)$ as input. For complex environments with multiple constraints, we propose to build this function by an arbitrary composition of conjunctions, disjunctions, and negations \cite{glotfelter_nonsmooth_2017}\cite{molnar2023composing}. To illustrate this, consider the sets $\mathcal{S}_i=L_0^+(s_i), i \in I$. The following relations hold for any $e$:
\begin{equation}\label{eq:boolean}
\begin{aligned}
\bigcup_{i\in I}\mathcal{S}_i &= \{x \in \mathcal{D}| \max_{i\in I} s_i(x,e)\geq 0\} \\
\bigcap_{i\in I}\mathcal{S}_i &= \{x\in \mathcal{D}| \min_{i\in I} s_i(x,e)\geq 0\}\\
\bar{\mathcal{S}}_i &=  \{x\in\mathcal{D}|-s_i(x,e)\geq 0 \}.
\end{aligned}
\end{equation}
Similarly, arbitrary combinations of unions and intersections can be captured by composing the $\max$ and $\min$ operators. 
\begin{example} \itshape
Consider the Dubins car dynamics $\dot{x}_1 =  V \cos(\psi), \dot{x}_2 = V \sin(\psi), \dot{\theta} = u$ where $(x_1, x_2)$ denote the position coordinates on the 2D plane, $\psi$ denotes the angle of the unicycle, $V$ is a constant speed, and $u$ is the angular velocity considered as control input. Suppose there are two circular obstacles in the environment which respectively give the following safety constraints:
\begin{equation*}
    \begin{aligned}
        \mathcal{S}_i &= \left \{ x \in \mathbb{R}^2 \mid \lVert x - o_i \rVert_i^2 \geq r_i^2 \right \} \ i=1,2,
    \end{aligned}
\end{equation*}
The parameter $e$ in this example is given by $e = [o_1^\top \ r_1 \ o_2^\top r_2]^\top \in \mathbb{R}^6$, and the safe set $\mathcal{X} = \mathcal{S}_1 \cap \mathcal{S}_2$ is represented as
\begin{equation*}
    \mathcal{X} =  \{ x \mid  c(x, e):= \min(  \lVert x - o_1 \rVert_2^2 - r_1^2,  \lVert x - o_2 \rVert_2^2 - r_2^2) 
 \geq 0 \}.
\end{equation*}
\end{example}
\medskip

The above composition procedure may not preserve differentiability, resulting in a non-smooth $c$ in the parameterization \eqref{eq:nonsmooth parameterization}. To address this, we can approximate these operators via the perspective of the LSE. These approximations for the $\max$ operator are as follows,
\begin{equation} \label{eq:smooth_bounds}
\begin{aligned}
&\mathrm{LSE}(s;\beta)-\tfrac{\log(|I|)}{\beta}\leq\max_{i \in I} s_i(x, e)\leq \mathrm{LSE}(s;\beta), \\
& -\mathrm{LSE}(-s;\beta)\leq \min_{i \in I} s_i(x, e) \leq -\mathrm{LSE}(-s;\beta) + \tfrac{\log(|I|)}{\beta},
\end{aligned}
\end{equation} becoming arbitrarily tight as $\beta \rightarrow + \infty$. Now for a given $c(\cdot,e)$ composed of $\max$ and $\min$ operators, we can construct a single smooth function $\underline{c}(\cdot,e)$ such that $\underline{c}(x,e) \leq c(x,e)$ for all $x \in \mathcal{D}$. By substituting $\underline{c}(x, e)$ for $c(x, e)$ in \eqref{eq:nonsmooth parameterization}, we obtain the following CBF parameterization,
\begin{equation} \label{eq:smooth parameterization}
    h_\theta(x, e) = \underline{c}(x, e) - \delta_\theta(x, e).
\end{equation}
It directly follows that $L_0^+({h}_{\theta}(\cdot, e)) \subseteq L_0^+(\underline{c}(\cdot, e)) \subseteq L_0^{+}(c(\cdot, e))=\mathcal{X}$ for any $e \in \mathbb{R}^{n_e}$.

\subsection{Pipeline}
Here, we summarize the training pipeline. For each environmental parameter $e$, the safe set is defined as $L_0^+(c(x, e))$. We first fix $\beta > 0$ and obtain a smooth lower bound $\underline{c}(x, e)$ of $c(\cdot)$ through~\eqref{eq:smooth_bounds}. Next, we parameterize the neural CBF operator as $h_\theta(x,e) = \underline{c}(x, e) - \delta_\theta(x, e)$. Finally, we optimize the training loss in~\eqref{eq:combined_loss} over the parameters $\theta$.
\section{Experiments}
\label{sec:experiments}
We evaluate the performance of the CBF operator learning framework and demonstrate its effectiveness as an adaptive safety filter through numerical experiments. In all experiments, the MLP $\delta_\theta$ consists of four hidden layers, each with 50 neurons and $\tanh$ activation functions. The output layer uses a softplus activation to ensure differentiability and non-negativity. The CBF operator is trained in JAX~\cite{jax2018github} using the ADAM optimizer~\cite{kingma2014adam} on a single T4 GPU.
\subsection{Double Integrator}
In the first example, we consider the double integrator dynamics:
\begin{equation}
    \begin{bmatrix} \dot{x} \\ \dot{v} \end{bmatrix} =   \begin{bmatrix} 0 & 1 \\ 0 & 0\end{bmatrix}   \begin{bmatrix} x \\ v \end{bmatrix} +  \begin{bmatrix} 0 \\ 1\end{bmatrix}u,
\end{equation}
where $u \in[-1,1]$, $x$ and $v$ denote the position and velocity, respectively. The safe set is restricted to the set $[0, 10]\times [-5,5]$ with two additional varying circular obstacles parameterized as $e = [r_1, x_{c1}, v_{c1}, r_2, x_{c2}, v_{c2}] $. The training data is composed of $10^3$ uniformly randomly sampled obstacle configurations. The radii were sampled from the interval $[1,2]$. For each configuration, $10^4$ state samples are drawn uniformly from $[-1, 11]\times [-6,6]$.

\begin{figure}[tb]
    \centering
    \includegraphics[width=0.45\textwidth]{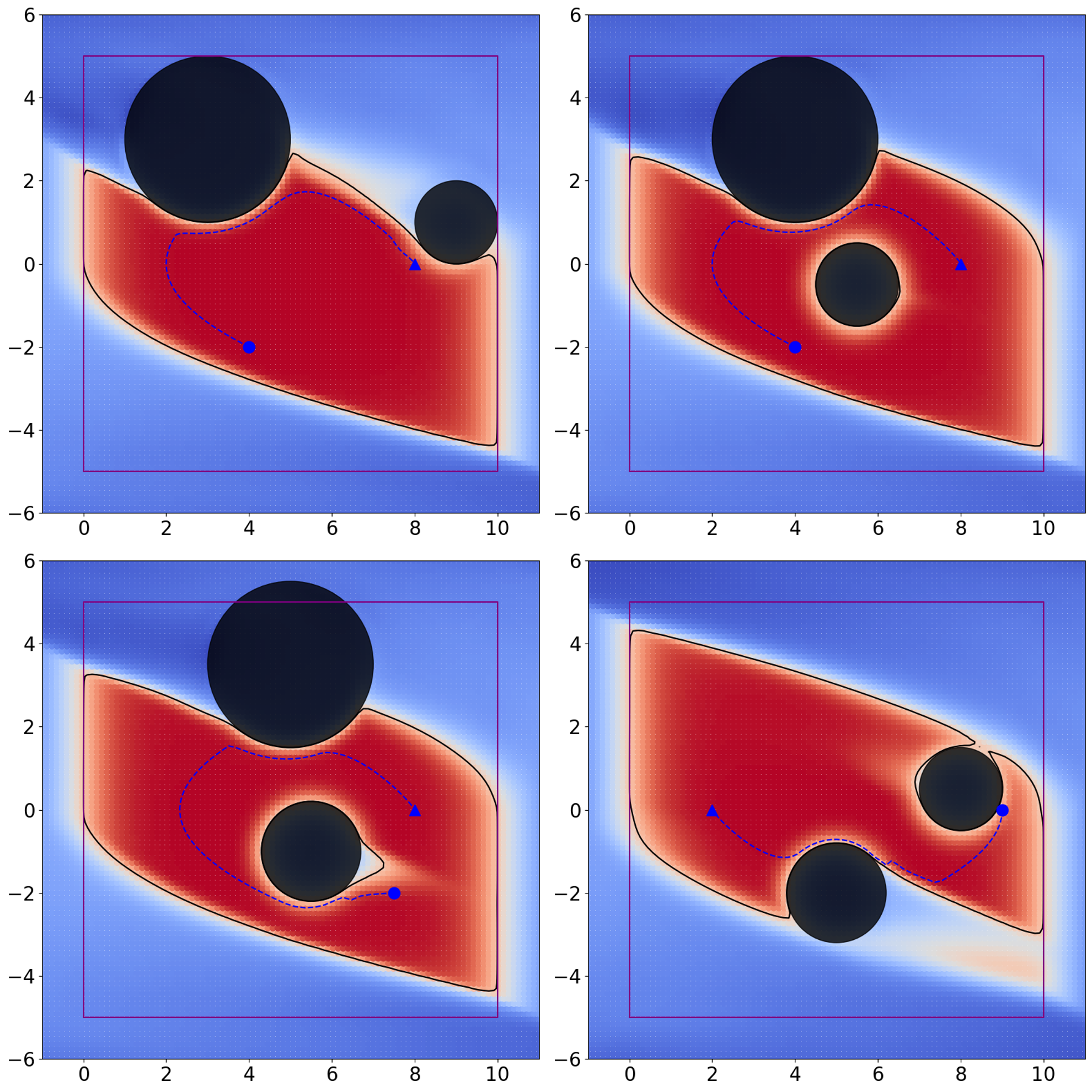}
    \caption{\bluecircle{} start state, \bluetriangle{} target state, \rule[0.5ex]{1em}{1pt} zero contour of the CBF instance $h_{\theta}(\cdot,e)$, horizontal-axis (position) and vertical-axis (velocity). The circular obstacles are marked black. For each environment, the CBF value $h_\theta(\cdot, e)$ is denoted by the color, with red denoting high values and blue denoting low values. }
    \label{fig:double_integrator}
\end{figure}

In Figure~\ref{fig:double_integrator}, we apply the learned CBF operator to filter a proportional-derivative reference controller. The underlying shape of $L_0^+(h_\theta(\cdot, e))$ is robust across environments and is close to the maximal control invariant set of the double integrator in obstacle-free environments. Furthermore, the closed-loop system trajectories under the adaptive safety filter demonstrate the efficacy of the learned CBF operator. 
  
\subsection{Unicycle} 
\begin{figure}[tb]
    \centering
    \includegraphics[width=0.5\textwidth]{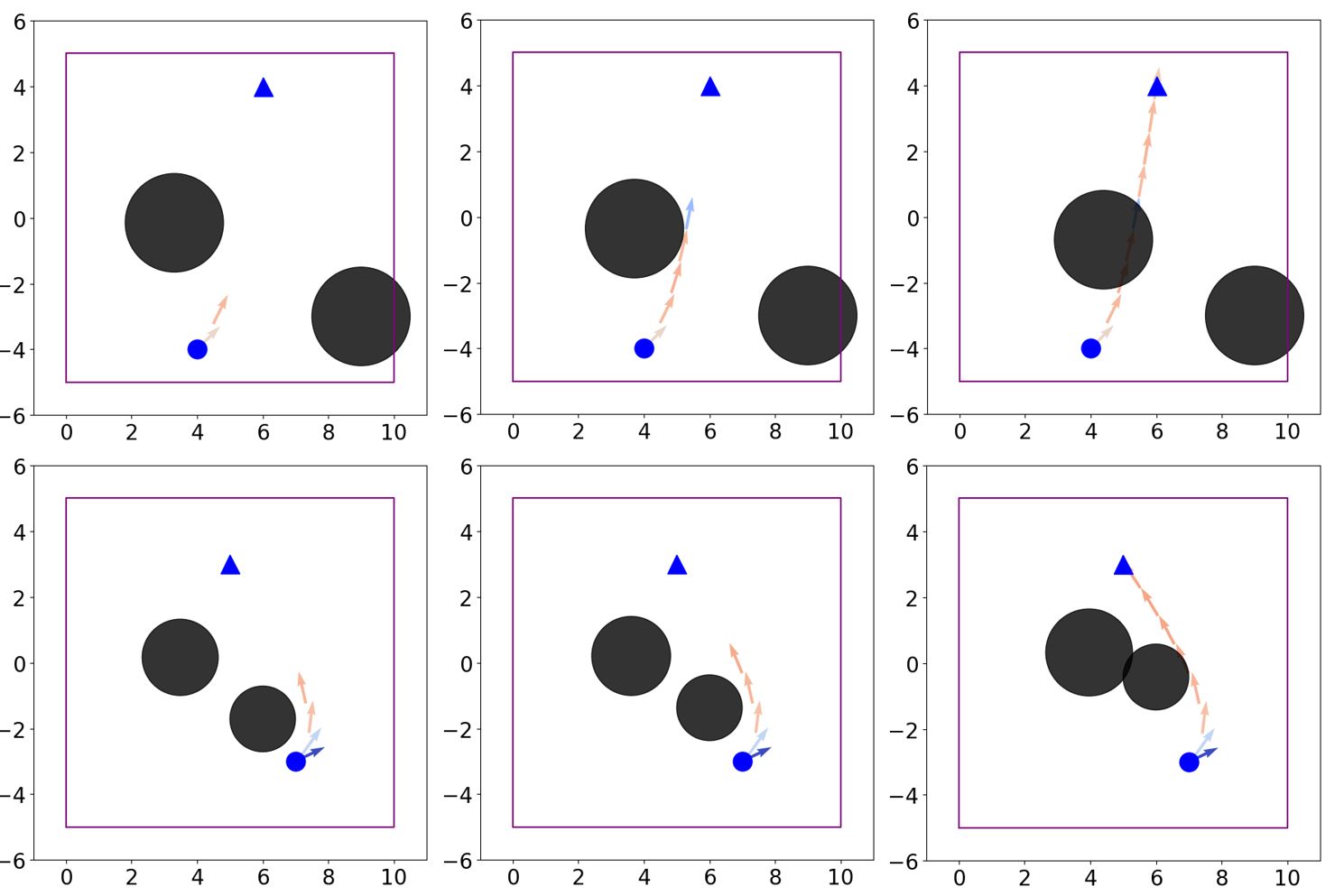}
    \caption{\bluecircle{}: start state. \bluetriangle{}: target state. The horizontal and vertical axes denote the position of the unicycle, and the arrow denotes its orientation. Top row: An obstacle moves from left to right trying to block the agent. Bottom row: the smaller obstacle chases the agent while the bigger obstacle moves slowly and grows over time. During motion, the CBF operator value changes in response to the moving obstacles. The heatmap on the arrows indicates the CBF values along the trajectory, with warmer colors denoting higher values. }
    \label{fig:unicycle}
\end{figure}

Next, we consider the unicycle system given by
\begin{equation}
    \begin{bmatrix} \dot{x} \\ \dot{y}\\ \dot{\psi} \end{bmatrix} = A(\psi) u, \; A(\psi) = \begin{bmatrix}\cos(\psi) & 0\\ \sin(\psi) & 0\\ 0 & 1\end{bmatrix},\;u = \begin{bmatrix} v \\ \omega \end{bmatrix},
\end{equation}
where $(x, y)$ denotes the position, $\psi$ denotes the orientation angle, $v$ is the velocity, and $\omega$ is the angular velocity. The controls are restricted to $\omega\in[-1,1]$ and $v\in[0.2,2]$. The safe set is given by  $(x, y, \psi) \in [0, 10]\times [-5,5]\times [-\infty, \infty]$ with two additional varying circular obstacles in $x,y$ (see Figure~\ref{fig:unicycle}). Importantly, the obstacles are moving which requires the system to apply different CBFs over time. As in the double integrator experiment, we sample $10^3$ environmental parameters but only $3\times 10^4$ states to demonstrate sample efficiency, as opposed to the 1 million states predicted by dimensional scaling $(10^4)^\frac{3}{2}$. The observed sample efficiency exceeds expectations based on simple scaling, highlighting the potential of CBF operator learning for higher-dimensional systems.

The reference controller for the unicycle is given by \cite{mestres_distributed_2024}:
\begin{equation}
    \mathrm{rerr}(x,y) = R(\psi)^{-1}\begin{bmatrix} x_f-x\\ y_f-y\end{bmatrix},\\ 
\end{equation}
$$d = \mathrm{atan2}(\mathrm{rerr}[1], \mathrm{rerr}[0]),\;\omega = k_{\omega}d,\; v = k_v\cos(d),$$
with $R(\psi)$ being a rotation matrix and $k_{\omega}, k_v>0$ arbitrary gains. We apply the learned CBF operator as the safety filter in Figure~\ref{fig:unicycle} with two different obstacle setups. In both of them, the unicycle system is able to make timely response and reach the target without collision. 

\section{Conclusion}
\label{sec:conclusion}
We proposed a deep operator learning framework to learn domain-adaptive control barrier functions that can adapt to novel environments without retraining. Leveraging domain randomization, the proposed framework learns the solution operator of a Hamilton-Jacobi PDE that maps the environment parameters to a control barrier function that approximates the maximal control invariant set. Our numerical experiments show the effectiveness of the learned CBF to ensure safe navigation in new environments. For future work, we will incorporate a falsification/verification module to refine the learned model based on counter-examples.

\bibliographystyle{ieeetr}
\bibliography{ref}

\end{document}